\documentclass[useAMS,usenatbib]{mn2e}
\usepackage{lscape,graphicx,natbib}

\title[A Polarimetric Study of the B{[e]} star HD 45677]{A Polarimetric Study of the B[e] star HD 45677}
\author[M. Patel, R. D. Oudmaijer, J. S. Vink, J. C. Mottram and B. Davies]{M. Patel$^{1}$, R. D. Oudmaijer$^{1}$, J. S. Vink$^{2,3}$, J. C. Mottram$^{1}$ and B. Davies$^{1}$ \\
$^{1}$School of Physics $\&$ Astronomy, University of Leeds, Woodhouse Lane, Leeds LS2 9JT, UK \\
$^{2}$Lennard-Jones Laboratory, Astrophysics, Keele University, Keele ST5 5BG, UK \\
$^{3}$Imperial College of Science, Technology and Medicine, Blackett Laboratory, Prince Consort Road, London SW7 2BZ, UK} 

\begin{document}
\date{}

\pagerange{\pageref{firstpage}--\pageref{lastpage}} \pubyear{2006}

\maketitle

\label{first page}

\begin{abstract}
We present new medium-resolution spectropolarimetric observations of
HD~45677 in the {\it B} and {\it R}-bands.  A change in polarisation
is detected across H$\alpha$, H$\beta$ and H$\gamma$ confirming that
the ionised region around the star is aspherical. (Q,U) points
associated with these emission lines occur away from the continuum,
defining a polarisation vector which points in the same direction for
each of the lines at an average intrinsic polarisation angle of
164$^\circ \pm 3^\circ$.

These data were combined with past photometric and polarimetric data
from the literature to investigate any variability. We find that
HD~45677 is both photometrically and polarimetrically variable and
that these changes are linked. We suggest that these variations may be
caused by an aspherical blowout and by deriving a least-squares fit to
the {\it B}-band polarimetric data in Q-U space, we show that the
blowout occurs at an intrinsic polarisation angle of 175$^\circ \pm
1^\circ$, along the same angle as the proposed geometry of the ionised
region.

\end{abstract}

 
\section{Introduction}
HD 45677 is a B2 V star \citep{Israelian_etal:1996} which was first
observed in the late 19th century. Spectroscopic observations revealed
forbidden emission lines leading to the star being used as a prototype
for the B[e] classification although, ironically, its nature is still
unclear (Lamers et al. 1998 and below). These lines suggested the
presence of circumstellar gas which implied that the star was evolved
as it was not near any star-forming regions. More recent evidence
however, such as the absence of CO (1-0), SiO and OH maser emission
and the presence of infall features, suggests that the object is not a
post-AGB star and may be young
\citep{deWinter_vandenAncker:1997,Grady_etal:1993,Muratorio_etal:2006}.

The geometry of the circumstellar material was proposed to be
disk-like by \cite{Swings:1973} from spectroscopic
observations. Polarimetric studies by \cite{Coyne_Vrba:1976} also
suggested that the observed polarisations were most likely due to
grains in a flattened disk. \cite{Sorrell:1989} modelled the
circumstellar environment using a spherical distribution of dust
grains and found that the data matched the model quite well. He
suggested that any discrepancies may be explained by an aspherical
shell. \cite{Schulte-Ladbeck_etal:1992} conducted a
spectropolarimetric survey of HD 45677. They found that the
polarisation spectra in the red and the {\it UV} significantly
differed from one another. Furthermore, after correcting for
interstellar polarisation, they observed a 90$^\circ$ rotation in
position angle from the {\it B}-band to the {\it UV}. They explained
these variations and the angle flip by proposing two sources of
polarisation - one that dominates in the red and one in the {\it
UV}. Furthermore, they suggest that the polarisation spectrum is
characteristic of a bipolar nebula. \cite{Oudmaijer_Drew:1999}
conducted medium-resolution spectropolarimetric observations of the
object in the optical and concluded that the geometry of the ionised
region is either disk-like or bipolar based on changes in polarisation
across H$\alpha$.

Spectropolarimetry is a powerful tool in probing the inner regions
around a star. In particular, the presence of line-effects in a
polarisation spectrum can indicate whether there is a non-spherical,
ionised region. Line-effects are changes in polarisation over spectral
features associated with emission. They occur as a result of the
emission line being formed in a volume far larger than the region
where electron-scattering is dominant. Subsequently, these emission
line photons interact with fewer electrons compared to the continuum
light and so experience fewer scatterings, resulting in a lower
polarisation at the line-centre.

The aim of this paper is to investigate the geometry of the
circumstellar material around HD~45677 by using various polarimetric
techniques.  To this end, we obtain new spectropolarimetric data and
collect past polarimetric and photometric data from the literature.

The paper is organised as follows, in Section 2 we discuss the
observational data on HD~45677. We explain the instrumental setup used
in the spectropolarimetric observations and give a brief summary of
the past polarimetric and photometric data. In Section 3, we review
the photometric results found by authors in the literature and we
present new spectropolarimetric data together with past polarimetric
data. In Section 4, we investigate the presence of line-effects in the
spectropolarimetric data and discuss the links between changes in the
photometric and polarimetric data. Finally, conclusions drawn from the
study are presented in Section 5.

\section{Observations}
\subsection{Literature Data}
The photometry of HD~45677 was comprehensively reviewed by
\cite{deWinter_vandenAncker:1997}. Therefore, we have used their data
and sources within their paper together with more recent data from the
AAVSO database \citep{Henden:2006} to build a history of the {\it
V}-band photometry of the star.  Table $\ref{T:PPhoto}$ gives a
summary of the photometric data collected on HD~45677.

\begin{table*}
\begin{minipage}{80mm}
\centering
\begin{tabular}{lccccccc}
\hline
\centering
Source & Time Interval \\
       & (JD 2440000+) \\
\hline
Feinstein et al. (1976) & 1686-2803 \\
Kilkenny et al. (1985) & 4934-5809 \\
Halbedel (1989) & 6391-7596 \\
De Winter $\&$ Van den Ancker (1997) & 3124-9359 \\
AAVSO & 2032-13706 \\
\hline 
\end{tabular}
\caption{A summary of the {\it V}-band photometry collected from the literature. In Column (2), we state the time interval over which each data set were taken.}
\label{T:PPhoto}
\end{minipage}
\end{table*}    

In addition, we have collected {\it B} and {\it R}-band polarimetric
measurements of HD~45677 from the literature. A list of these
observations together with their results can be found in Table
$\ref{T:HD45677_past}$. Errors were taken directly from the literature.
When errors in position angle were not available, they were calculated
using
 \begin{equation}
 \delta\theta = 28^\circ\, \times \, (\frac{\delta P}{P})
\end{equation}

\subsection{Spectropolarimetric Observations}
The linear spectropolarimetric data were taken on the 28th and 29th
September 2004 using the ISIS spectrograph mounted on the 4.2m William
Herschel Telescope (WHT), La Palma. The seeing on both nights was
$<1.7''$. The star was observed twice in the {\it B}-band using a
4096x2048 pixel EEV12 CCD detector with a R1200B grating, which
yielded a spectral coverage of 4295-4955\AA$ $ and a spectral
resolution of 51 km/s around H$\beta$. For the {\it R}-band
observation, we used a MARCONI2 CCD detector with the R1200R grating
giving a spectral range of 6150-6815\AA$ $ and a spectral
resolution of 34 km/s around H$\alpha$. For all of the images a slit
width of 1 arcsec was used.

The instrumental setup consisted of a calcite block which splits the
light into two perpendicularly polarized beams (the o and e rays) and
a half-wave plate used to rotate the polarisation of the incoming
light.  There were two extra holes in the dekker mask through which
the sky was simultaneously observed, so for each image a total of six
spectra were obtained. One complete polarisation set consisted of four
(two for the sky and two for the object) measurements taken at
0$^\circ$ and 45$^\circ$ (to measure Stokes Q) and 22.5$^\circ$ and
67.5$^\circ$ (to measure Stokes U).

Data reduction was carried out using the Figaro software maintained by
Starlink, and consisted of bias subtraction, cosmic ray removal, bad
pixel correction, spectrum straightening and flat fielding. Wavelength
calibration used several spectra (obtained by observing a Copper-Argon
lamp) taken throughout the observing run. To obtain the Stokes
parameters, each polarisation set was fed into the Time-Series
Polarimetry (TSP) package, also maintained by Starlink. The data was
position angle-calibrated using a set of polarisation standard stars
observed during the run. Data analysis was carried out using the
Starlink package POLMAP.

In principle, polarisation data is only limited by photon statistics. In
practice, however, systematic errors such as the effects of 
scattered light and instrumental polarisation become dominant. The 
observations of two zero-polarisation objects show that the  
instrumental polarisation is around 0.10-0.15$\%$.

During our reduction we found large changes in polarisation across the
centre of the Balmer lines, whose features were narrower than the
spectral resolution. These are most likely related to a slightly
different focus in the o and e rays - we measured a resolution
difference of 0.1 pixels between these rays. This difference affects
the data considerably where barely-resolved spectral features are
present, producing narrow line-effects over the central absorption
components of H$\alpha$, H$\beta$ and H$\gamma$. These areas were
removed and were not used during the analysis of the star. 

\section{Results}
\begin{figure}
\centering
\includegraphics[width=80mm, height=100mm]{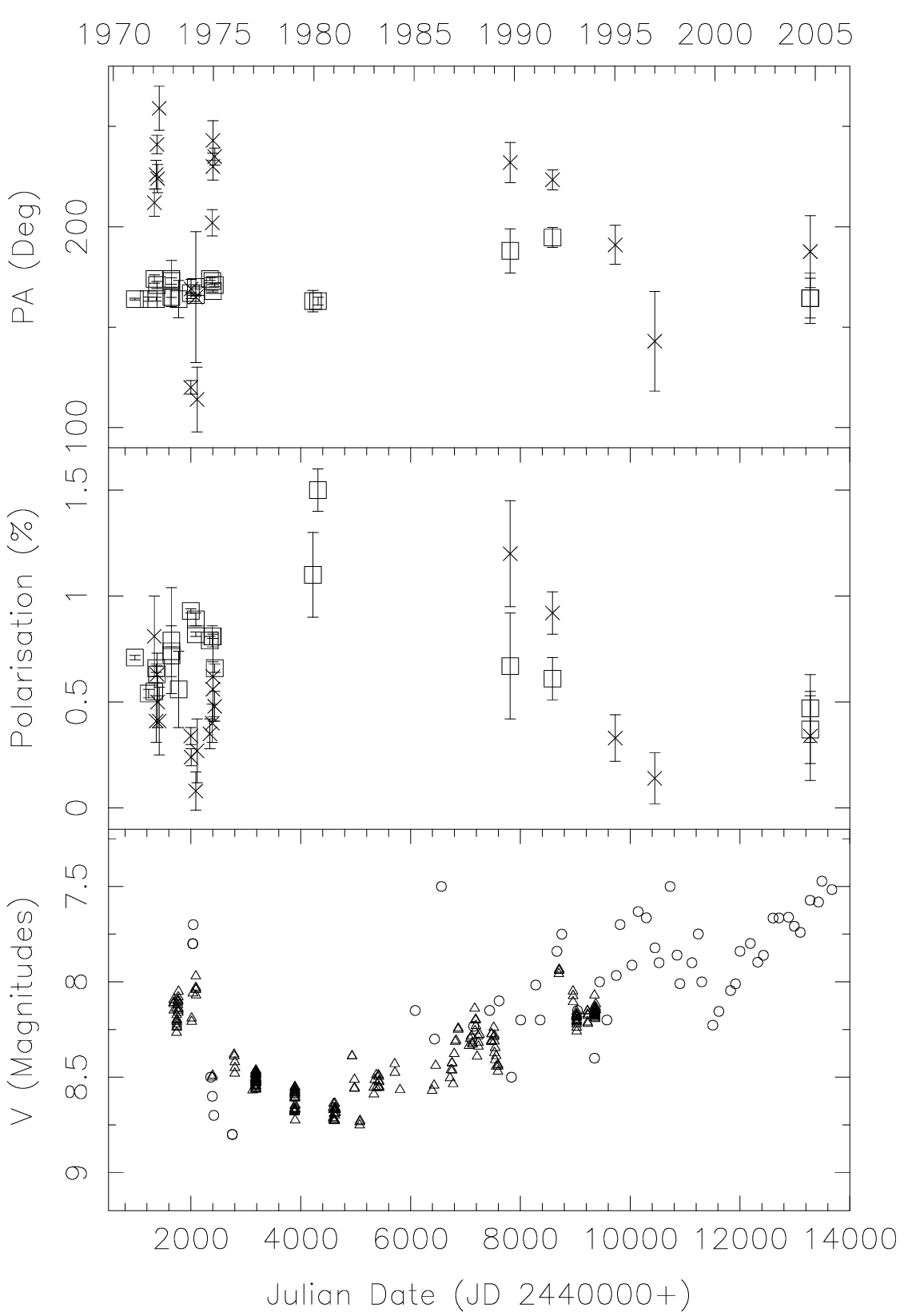}
\caption{The variability of HD~45677 in the {\it B} and {\it R}-bands over the last 30 years. The top and middle panels plot the position angle and polarimetric variability, respectively. Squares represent {\it B}-band data and crosses indicate {\it R}-band data. Both {\it B} and {\it R}-band data show significant variations in the polarisation. The lower panel shows {\it V}-band photometry for HD~45677 taken by de Winter $\&$ van den Ancker (1997), Feinstein et al. (1976), Kilkenny et al. (1985) $\&$ Halbedel (1989) [triangles] $\&$ AAVSO [circles]. The data has been grouped so that multiple observations within a short period of time are averaged. The errors associated with the photometry are less than 0.025 mag.}
\label{F:HD45677_var}
\end{figure}
Our results are threefold, in the first section ($\S$3.1) we provide a brief
review of the photometric data collected on HD~45677. In the next
section ($\S$3.2), we present new spectropolarimetric data taken in the {\it B}
and {\it R}-bands. In our final section ($\S$3.3), we combine these new
measurements with past polarimetric observations taken within the same
passband.

\subsection{Past Photometric Data}

We are mainly interested in the {\it V}-band photometric data from
1972 to 2005 as this is the same period over which polarimetric data
has been obtained.  The {\it V}-band variations of HD~45677 are shown
in the bottom panel of Fig. $\ref{F:HD45677_var}$.  The star decreased
in brightness from 1975 to 1980, but became brighter later on,
resulting in a minimum in visual brightness in the early 1980's. De
Winter $\&$ van den Ancker (1997) stated that a wide binary may cause the
photometric variations, however the asymmetry of the variations and
the absence of periodicity in the radial velocity measurements suggests
that this may not be the case. Alternatively, they suggest that a
blowout of material could have occurred in the 1950s resulting in the
gradual obscuration of the star by large dust grains. The presence of
a minimum and the eventual brightening is then explained by the destruction
of these dust grains.

\subsection{Spectropolarimetry}
In our analysis, we shall only take account of the data around the
Balmer lines, ignoring any line-effects which occur over barely
resolved line components for reasons explained in Section 2.2.

\begin{figure*}
\centering
\includegraphics[width=50mm, height=70mm]{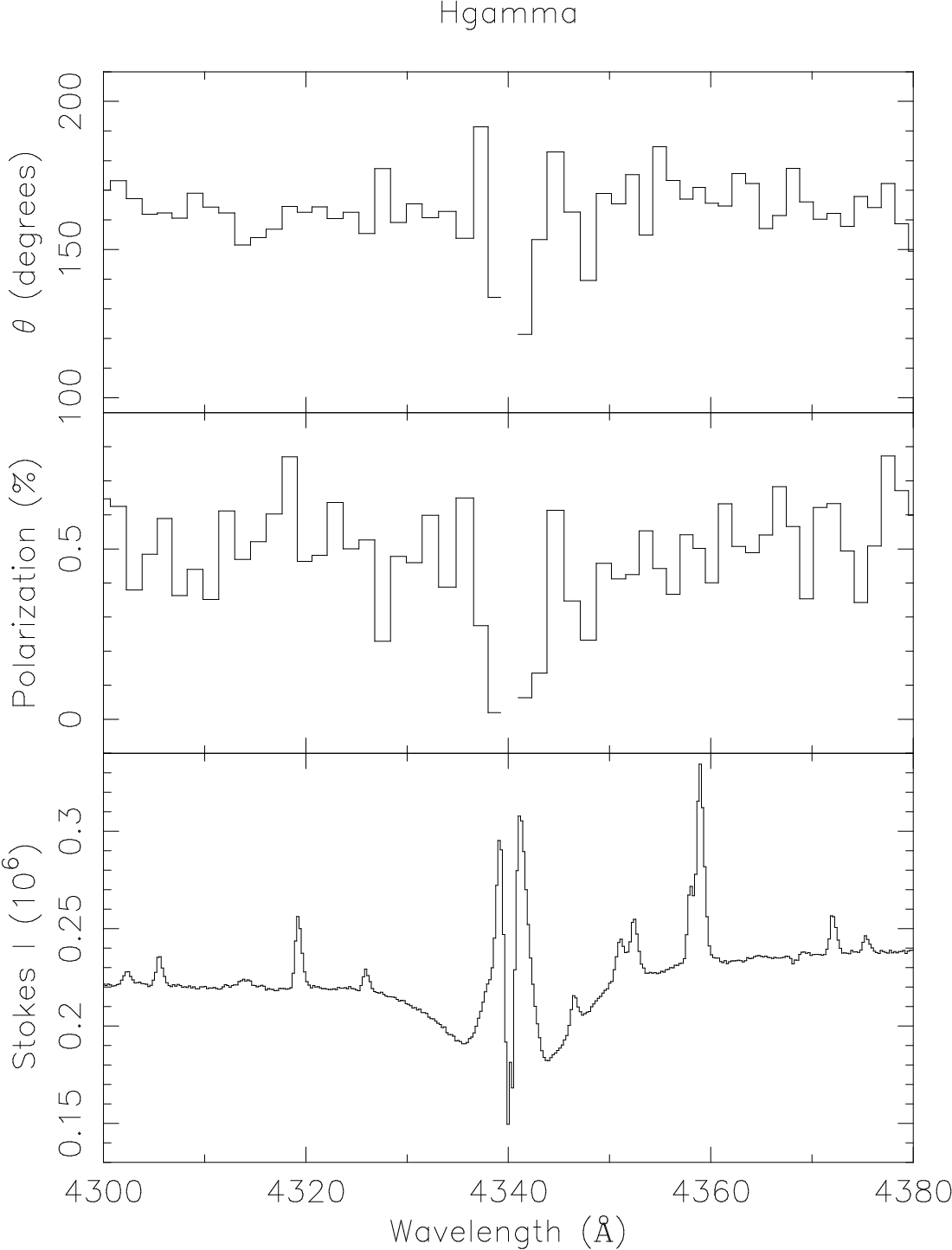}
\includegraphics[width=50mm, height=70mm]{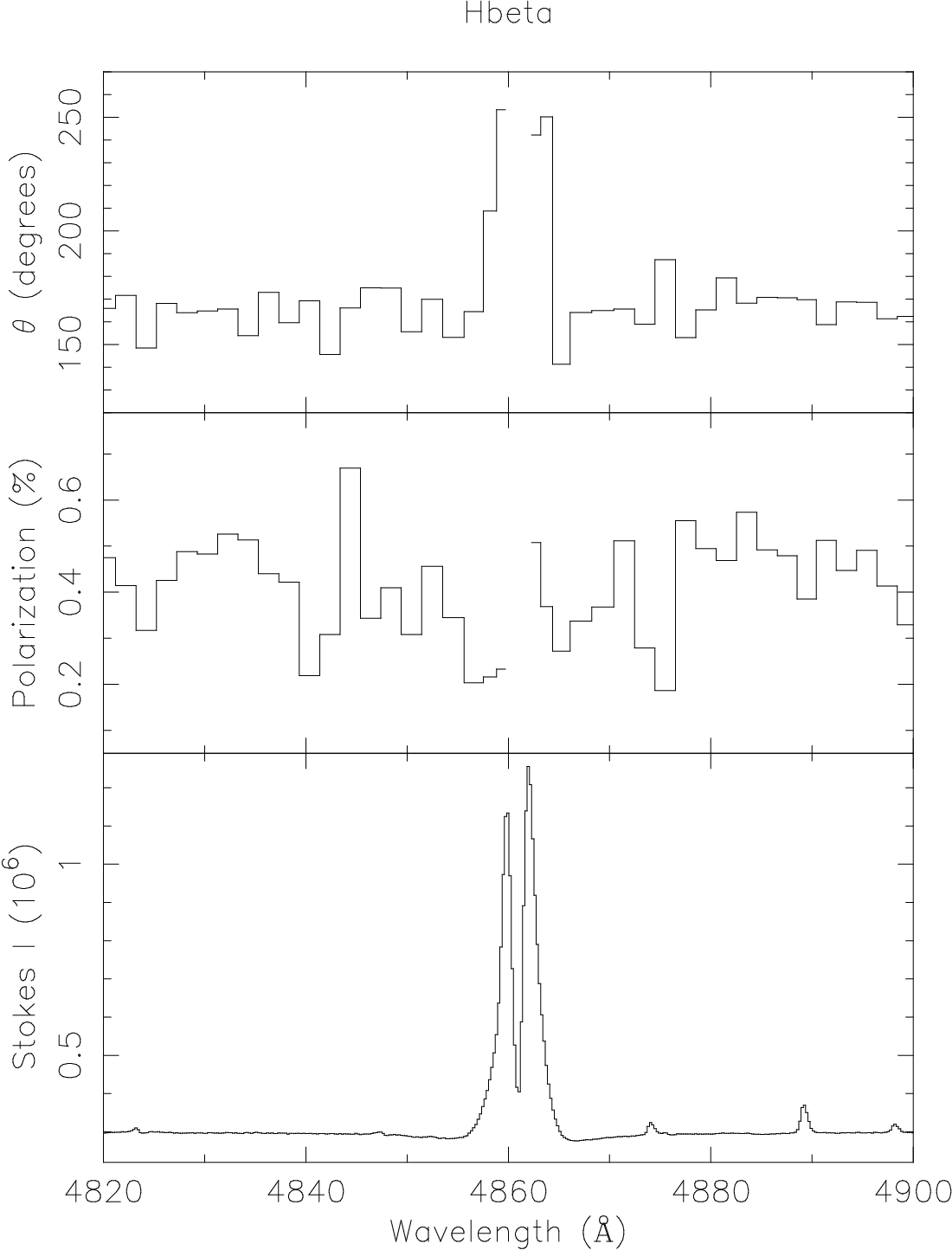}
\includegraphics[width=50mm, height=70mm]{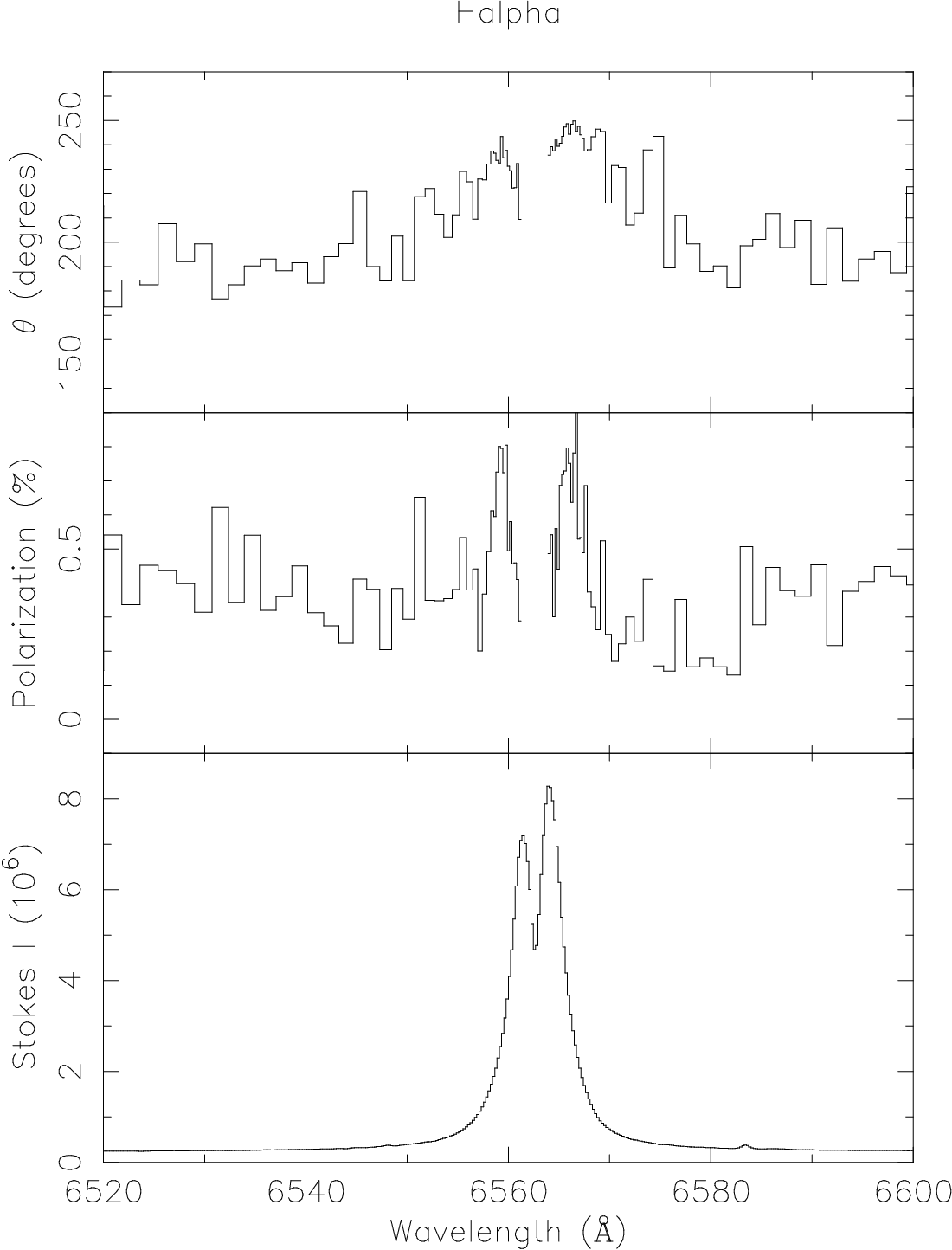}
\vfill
\vspace{2mm}
\includegraphics[width=30mm, height=30mm]{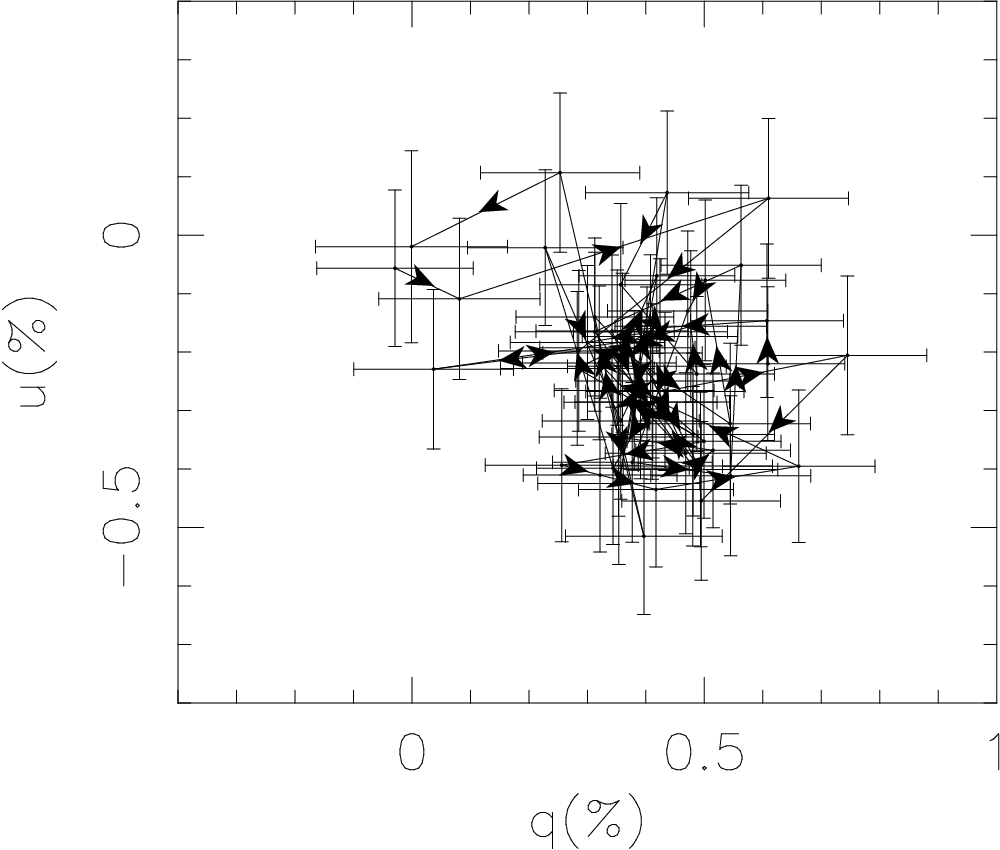}
\hspace{20mm}
\includegraphics[width=30mm, height=30mm]{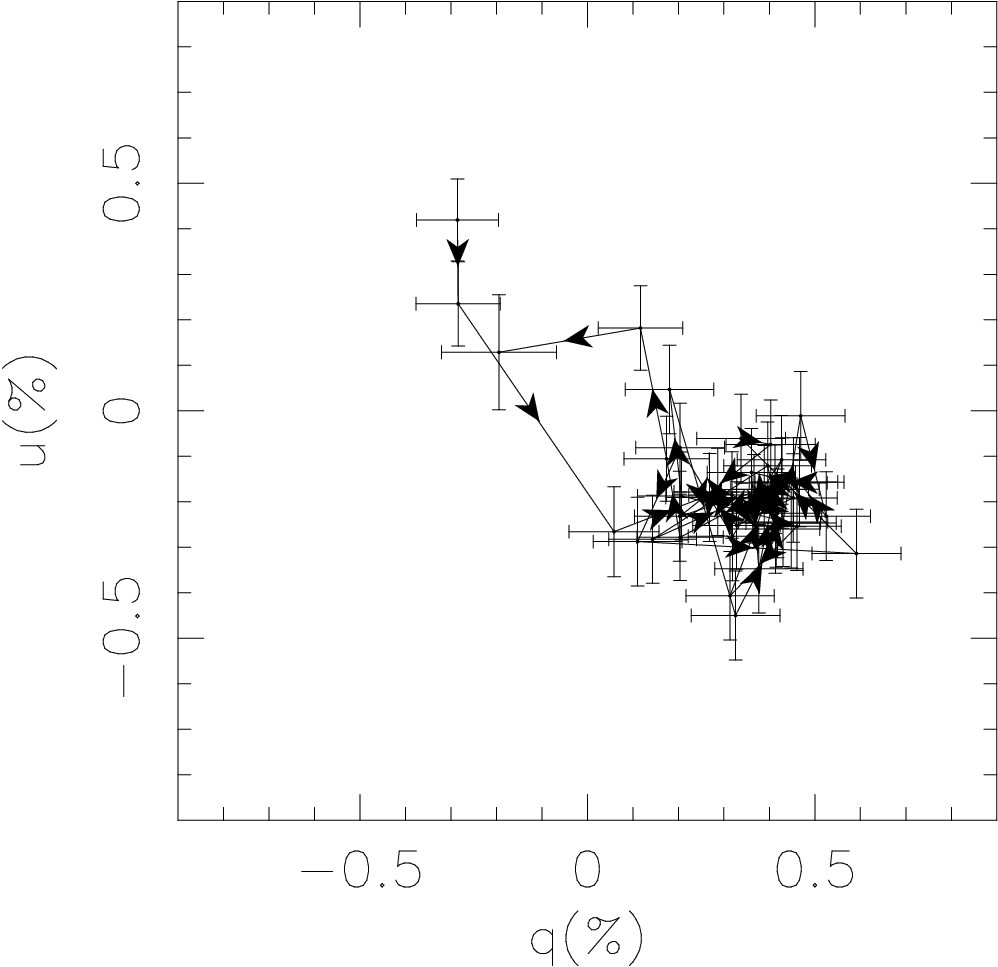}
\hspace{20mm}
\includegraphics[width=30mm, height=30mm]{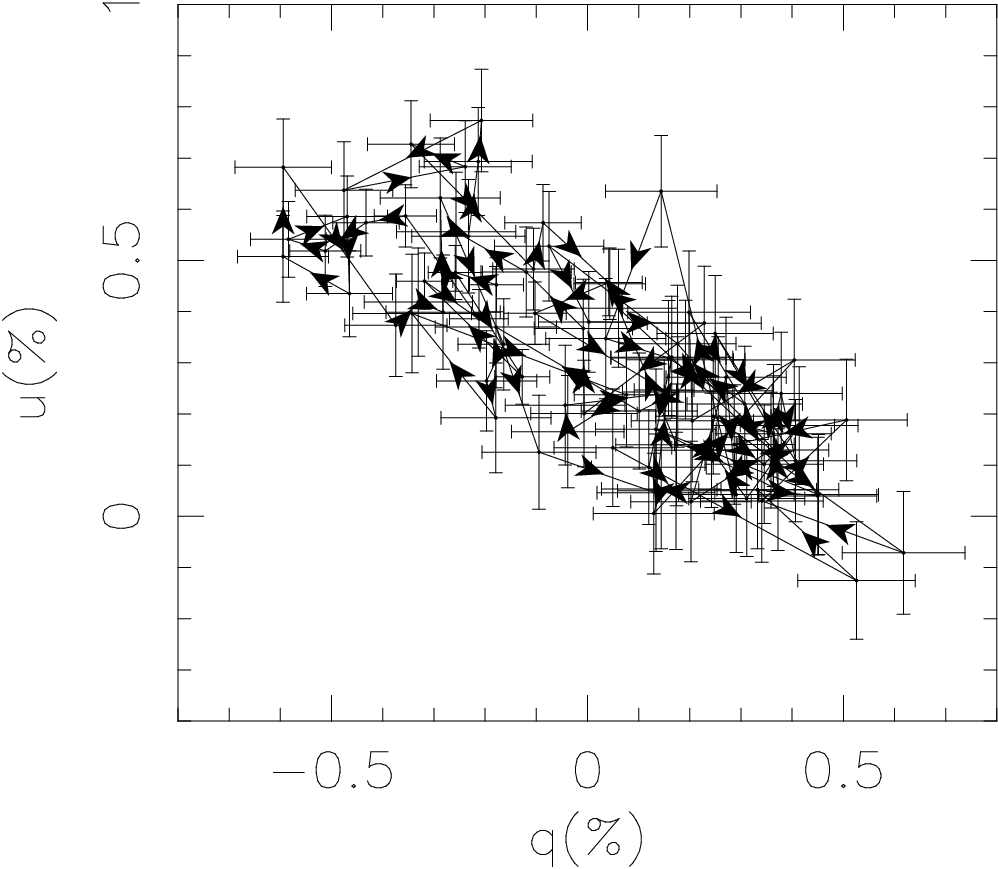}
\caption{The polarisation spectra and corresponding (Q,U) plots of HD~45677 around H$\gamma$, H$\beta$ and H$\alpha$ (left to right) taken on 29/09/04. Plotted in the lowest triplot panel is the Stokes I spectrum, the polarisation and position angle spectra are plotted in the middle and upper panels, respectively. The data has been rebinned such that the 1$\sigma$ error in polarisation corresponds to 0.14$\%$ for H$\gamma$, 0.1$\%$ for H$\beta$ and 0.12$\%$ for H$\alpha$, as calculated from photon statistics. The plot below graphs the normalised Stokes parameters u = ($\frac{U}{I}$) and q = ($\frac{Q}{I}$) at the same bins as their corresponding triplots. The central part of the spectrum has been removed for reasons stated in the text. Line-effects can be seen across all three lines on the 29th, with a depolarisation across H$\gamma$, no significant polarimetric effect at H$\beta$ (but a rotation in angle) and an enhancement in polarisation across H$\alpha$. The (Q,U) points associated with the Balmer lines occur away from the continuum and at similar angles to one another, defining an average intrinsic polarisation angle of 164$^\circ \pm 3^\circ$. }
\label{F:HD45677_le}
\end{figure*}

Fig. $\ref{F:HD45677_le}$ shows the polarisation spectrum of HD~45677
around H$\gamma$, H$\beta$ and H$\alpha$ taken on 29/09/04. {\it
B}-band data taken on the 28th has a lower S/N and therefore is less
clear. However, similar profiles in PA and polarisation can be seen
across H$\beta$. The different panels in each plot display the
position angle, polarisation and Stokes I parameter (top to
bottom). The lower diagrams are the same data set represented in Q-U
space.  We see line-effects across all three Balmer lines. H$\gamma$
is a double-peaked line whose emission peaks are just above the
continuum. Across the emission peaks, we see a
depolarisation. H$\beta$ is a fairly strong double-peaked emission
line, but no significant changes in polarisation can be seen across
the line, although a rotation in angle occurs. H$\alpha$ is a strong,
broad, double-peaked emission line. There is a change in both the
polarisation and PA across the entire line, extending out to the
broad wings. The presence of these line-effects indicates that the
electron-scattering region around HD~45677 is aspherical. 

We note, however, that the spectropolarimetric line profiles can be
affected by intervening polarisation due to circumstellar and
interstellar dust, as the observed polarisation is due to a vector sum
of these components with the polarisation due to electrons. This can
turn an otherwise straightforward depolarisation into more complex
profiles (see e.g. the case of HD 87643, Oudmaijer et al. 1998).  If
we represent the data in (Q,U) space, however, the intervening
polarisation contributes only a constant (Q,U) vector, leaving the
intrinsic shape of the line-effect unaffected
(Fig. \ref{F:HD45677_le}). The (Q,U) points associated with the
H$\alpha$, H$\beta$ and H$\gamma$ emission lines are located away from
the continuum and significantly, do so at the same angle.  

The (Q,U) vectors from the continuum to the line-centres define the
electron polarisation vector, and the angle it makes with the Q axis
is a measure of the intrinsic polarisation angle. This can be directly
related to the geometry of the ionised region using
\begin{equation}
 \theta = 0.5\, \times \, \arctan\,(\frac{\Delta U}{\Delta Q})
\end{equation}
where $\Delta$U and $\Delta$Q are the difference between line and
continuum values of U and Q, respectively.

$\newline$ Since the line-centres had to be removed from our data due
to resolution effects, we are unable to calculate the magnitude of the
electron polarisation vector. However, we are still able to measure
the intrinsic polarisation angle. The direction of the electron
polarisation vector associated with the line-effects across all three
Balmer lines are similar and have an average intrinsic polarisation
angle of 164$^\circ$ $\pm$ 3$^\circ$.

We can estimate the interstellar polarisation component of HD~45677 by
assuming that H$\alpha$ is formed far from the electron-scattering
region, and therefore that the polarisation at the line-centre is due
to circumstellar and interstellar dust alone.  Again, since the Balmer
line-centres have been removed we cannot calculate the interstellar
polarisation from our data. And so, we will use data from the
literature.

\begin{figure*}
\centering
\includegraphics[angle=-90,scale=0.6]{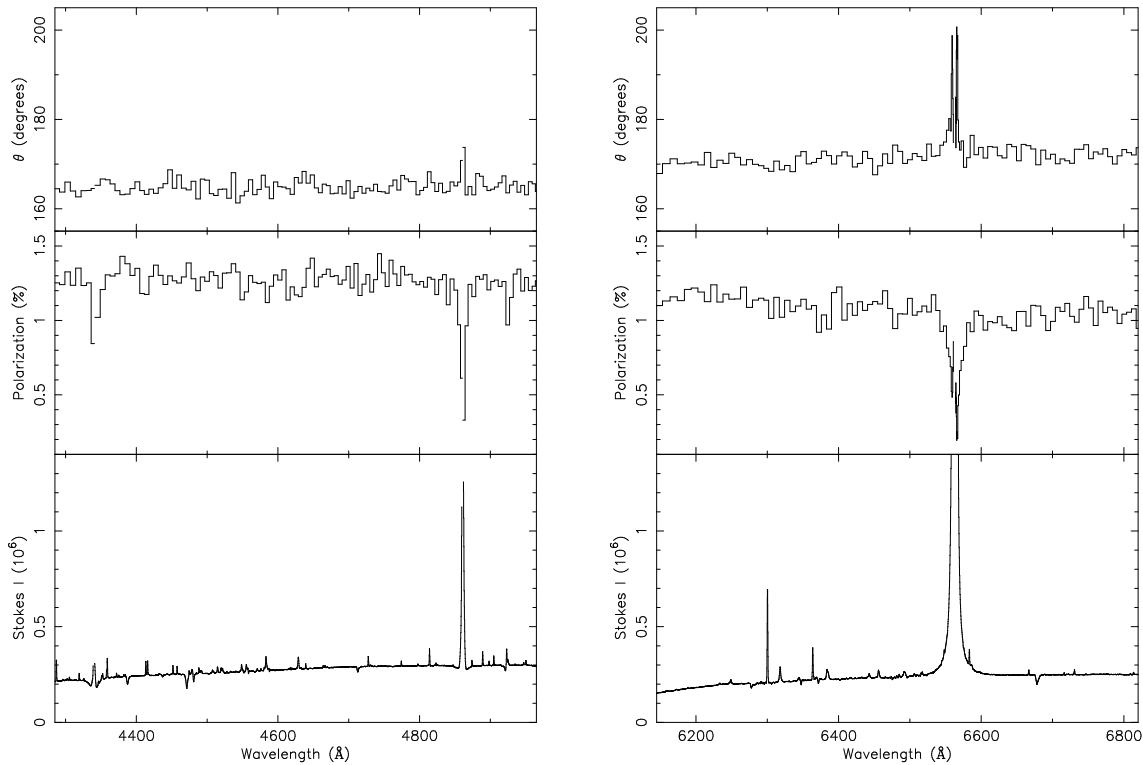}
\caption{The intrinsic polarisation spectrum of HD~45677 taken on 29/09/04 over the {\it B} and {\it R}-bands. The data are rebinned such that the 1$\sigma$ error in the polarisation  corresponds to 0.06$\%$ as calculated from photon statistics. The intrinsic spectrum was produced by removing an estimation of the ISP, using the line-centre polarisation at H$\alpha$ measured by Oudmaijer $\&$ Drew (1999), from the polarisation data collected in 2004. The Balmer lines all show depolarisations indicating that the enhancement initially observed was due to foreground polarisation.}
\label{F:HD45677_whole}
\end{figure*}

\cite{Oudmaijer_Drew:1999} looked at the polarisation spectrum around
H$\alpha$ twice within a one year period and found intrinsic angles of
163$^\circ$ $\pm$ 3$^\circ$ in 1995 and 168$^\circ$ $\pm$ 3$^\circ$ in
1996 and a similar H$\alpha$ line-centre polarisation at each
epoch. The consistency of these results implies that the line-centre
polarisation at H$\alpha$ remains the same. Therefore, we can use
measurements taken by Oudmaijer $\&$ Drew to infer an interstellar
polarisation and obtain an intrinsic spectrum.  The authors found the
H$\alpha$ line-centre polarisation to be 0.8$\%$ at an angle of
75$^\circ$.  Using these results and adopting a $\lambda_{max}$ of
5000\AA$ $ for this area of the sky \citep{Serkowski_etal:1975}, we
calculated a Serkowski-law with the parameters p$_{max}$ = 0.87$\%$,
$\theta$ = 75$^\circ$ and removed this ISP component from the data to
produce the intrinsic polarisation spectrum shown in
Fig. $\ref{F:HD45677_whole}$.

Across all of the Balmer lines, we now observe depolarisations. As the
polarisation goes to zero, the relative error increases and so, any
angle changes associated with the depolarisation are highly
uncertain. We therefore attribute the two large rotations seen across
H$\alpha$ as being due to such effects. The presence of
depolarisations across the Balmer lines in the intrinsic spectrum of
HD~45677 suggests that H$\alpha$, H$\beta$ and H$\gamma$ are formed
far from the electron-scattering region. 

\begin {table*}
\begin {minipage}{165mm}
\centering
\begin {tabular}{lcccc}
\hline
Paper & Date Observed & {\it B}/{\it R} band & $\%$P & P.A. (Deg) \\
\hline
Coyne $\&$ Vrba (1976) & 29/01/71 & {\it B} & 0.71 $\pm$ 0.01 & 164 $\pm$ 0.4 \\	
 & 08/10/71 & {\it B} & 0.54 $\pm$ 0.02 & 164 $\pm$ 1.1	\\
 & 19/01/72 & {\it B} & 0.55 $\pm$ 0.04 & 174 $\pm$ 2.1	\\
 & 21/02/72 & {\it B} & 0.66 $\pm$ 0.02 & 164 $\pm$ 0.9	\\
 & 07/03/72 & {\it B} & 0.63 $\pm$ 0.04 & 171 $\pm$ 1.8	\\
 & 22/11/72 & {\it B} & 0.74 $\pm$ 0.12 & 165 $\pm$ 4.7	\\
 & 23/11/72 & {\it B} & 0.79 $\pm$ 0.25 & 174 $\pm$ 9.2	\\
 & 24/11/72 & {\it B} & 0.72 $\pm$ 0.04 & 173 $\pm$ 1.6	\\
 & 04/04/73 & {\it B} & 0.56 $\pm$ 0.18 & 164 $\pm$ 9.3	\\
 & 12/11/73 & {\it B} & 0.93 $\pm$ 0.01 & 167 $\pm$ 0.3	\\
 & 11/02/74 & {\it B} & 0.89 $\pm$ 0.03 & 170 $\pm$ 1.0	\\
 & 12/02/74 & {\it B} & 0.82 $\pm$ 0.01 & 166 $\pm$ 0.4	\\
 & 24/10/74 & {\it B} & 0.79 $\pm$ 0.02 & 174 $\pm$ 0.7	\\
 & 08/12/74 & {\it B} & 0.81 $\pm$ 0.05 & 173 $\pm$ 1.8	\\
 & 18/12/74 & {\it B} & 0.81 $\pm$ 0.01 & 168 $\pm$ 0.4	\\
 & 21/01/75 & {\it B} & 0.66 $\pm$ 0.02 & 171 $\pm$ 0.9	\\
Barbier $\&$ Swings (1982)  & 12/79      & {\it B} & 1.10 $\pm$ 0.20 & 173.0 $\pm$ 5.3 \\
                            & 03/80      & {\it B} & 1.50 $\pm$ 0.10 & 173.0 $\pm$ 1.9 \\
Gnedin et al. (1992)        & 10/89      & {\it B} & 0.67 $\pm$ 0.25 & 8.0 $\pm$ 11.0 \\
Schulte-Ladbeck et al.(1992) & 12/90     & {\it B} & 0.61 $\pm$ 0.10 & 14.7 $\pm$ 5.0 \\
This paper                  & 28/09/04   & {\it B} & 0.37 $\pm$ 0.16 & 164.0 $\pm$ 12.1 \\
                            & 29/09/04   & {\it B} & 0.47 $\pm$ 0.16 & 165.0 $\pm$ 9.5 \\
Coyne $\&$ Vrba (1976) & 17/01/72 & {\it R} & 0.81 $\pm$ 0.19 & 212 $\pm$ 6.8 \\
 & 21/02/72 & {\it R} & 0.41 $\pm$ 0.10 & 226 $\pm$ 7.1 \\
 & 08/03/72 & {\it R} & 0.63 $\pm$ 0.10 & 241 $\pm$ 4.6 \\
 & 18/03/72 & {\it R} & 0.50 $\pm$ 0.12 & 224 $\pm$ 7.0 \\
 & 16/04/72 & {\it R} & 0.41 $\pm$ 0.16 & 259 $\pm$ 11 \\
 & 12/11/73 & {\it R} & 0.34 $\pm$ 0.04 & 120 $\pm$ 3.4 \\
 & 17/11/73 & {\it R} & 0.24 $\pm$ 0.04 & 169 $\pm$ 4.8 \\
 & 11/02/74 & {\it R} & 0.08 $\pm$ 0.09 & 165 $\pm$ 32.6 \\
 & 06/03/74 & {\it R} & 0.27 $\pm$ 0.15 & 114 $\pm$ 16.1 \\
 & 25/10/74 & {\it R} & 0.35 $\pm$ 0.07 & \\
 & 08/12/74 & {\it R} & 0.40 $\pm$ 0.09 & 202 $\pm$ 6.5 \\
 & 19/12/74 & {\it R} & 0.56 $\pm$ 0.13 & 230 $\pm$ 6.7 \\
 & 20/12/74 & {\it R} & 0.62 $\pm$ 0.21 & 243 $\pm$ 9.8 \\
 & 21/01/75 & {\it R} & 0.48 $\pm$ 0.07 & 235 $\pm$ 4.2 \\
Gnedin et al. (1992)        & 10/89      & {\it R} & 1.20 $\pm$ 0.25 & 52.0 $\pm$ 10.0 \\
Schulte-Ladbeck et al.(1992) & 12/90     & {\it R} & 0.92 $\pm$ 0.10 & 43.4 $\pm$ 5.0 \\
Oudmaijer $\&$ Drew (1999)  & 10/01/95   & {\it R} & 0.33 $\pm$ 0.11 & 11.0 $\pm$ 9.7 \\
                            & 29/12/96   & {\it R} & 0.14 $\pm$ 0.12 & 143.0 $\pm$ 24.8 \\
This paper                  & 29/09/04   & {\it R} & 0.34 $\pm$ 0.22 & 8.0 $\pm$ 18.1  \\
\hline 
\end {tabular} 
\caption{Past and present polarisation observations of HD~45677. Column (1) gives the authors who have obtained the measurements. Where no errors in angle were given, errors were calculated using the method described in Section 2.1. Measurements from new spectropolarimetric data (i.e. this paper) were obtained by measuring average polarisations and position angles between 4400-4800\AA$ $ in the {\it B}-band and between 6200-6500\AA $ $  and 6600-6800\AA $ $ in the {\it R}-band.} 
\label{T:HD45677_past}
\end {minipage}
\end {table*}

\subsection{Broad Band Polarisation}

A number of authors have conducted polarimetric studies of HD~45677,
Table $\ref{T:HD45677_past}$ gives a summary of the polarimetric
results found in the literature in the {\it B} and {\it R}-bands.                           
From past measurements and our new spectropolarimetric data (Table
$\ref{T:HD45677_past}$), we find that HD~45677 experiences long-term
polarimetric variations in both {\it B} and {\it R}-bands.  
The middle panel of Fig. $\ref{F:HD45677_var}$ shows that the {\it
B}-band polarisation peaks between 1980 and 1985, but decreases
from then onwards.  {\it R}-band polarimetric data suggests a similar
history but without any measurements from 1976 to 1988 this cannot be
confirmed. In the {\it R}-band, a rotation of 90$^\circ$ occurs within
a two year period, while angle changes in the {\it B}-band do not
exceed 30$^\circ$ over 32 years.

Comparing the polarimetric and photometric observations, we find that
as the observed {\it B}-band polarisation increases (from 1975 to
1980) the star becomes fainter and as the polarisation decreases (from
1982 to 1995) the star becomes brighter. Furthermore, the minimum in
visual brightness corresponds to a peak in polarisation. These changes
are similar to those of UXOR variables, where surrounding dust clouds
cause simultaneous changes in polarisation and photometry. However, de
Winter $\&$ van den Ancker investigated the {\it B}-{\it V} photometry
of HD~45677 and show that the colour of the object continues to become
redder even when the star is at minimum visual brightness. This is
inconsistent with UXOR variations.

Let us now discuss the observation that the polarisation angles are
more variable in the {\it R}-band than the {\it B}-band. In general,
the {\it R}-band data has a lower continuum polarisation than in the
{\it B}-band. Since the polarisation of HD~45677 is already low,
similar absolute errors are more significant in the {\it R}-band than
those in the {\it B}-band. This affects the position angle
measurements more than the corresponding polarisation measurements, as
the uncertainty in the determination of the PA, which results from a
ratio, will be substantial in the case of large errors bars in Q and
U. And so, the errors in PA in the {\it R}-band are often greater than
those which can be calculated using Eqn 1.
 
When we ignore all position angles for observations with a
polarisation smaller than 3$\sigma$, we find that most of the {\it
R}-band data is removed. This means that the quality of the {\it
R}-band data is inadequate to investigate whether the PA rotates more
in the {\it R}-band than in the {\it B}-band.

In Fig. \ref{F:HD45677_QU} we investigate the distribution of the
polarimetric data in Q-U space. We do not use the {\it R}-band data as
significant errors in Q and U will result in a large uncertainty in
the (Q,U) vectors between points, producing a large uncertainty in the
observed distribution. We find that the {\it B}-band data is well
fitted using a least-squares fit. This implies that the relationship
between Q and U is linear and the geometry associated with the
production of the {\it B}-band polarisation remains at the same angle
over time.  The gradient of the fit is related to the geometry of the
circumstellar material by Eqn. 2. We find an intrinsic PA of
175$^\circ \pm 1^\circ$. This is remarkably close to the value derived
for the electron-scattering region.

\section{Discussion}
We have presented medium-resolution spectropolarimetric observations
of HD~45677. The presence of line-effects within the data set suggests
that the geometry of the ionised region around the object is
aspherical.  Using H$\alpha$ line-centre polarisation measurements
taken by Oudmaijer $\&$ Drew (1999) we calculated a Serkowski-law with the
parameters p$_{max}$ = 0.87$\%$, $\theta$ = 75$^\circ$ at a
$\lambda_{max}$=5000\AA $ $ in order to arrive at the intrinsic
polarisation due to electron scattering alone.

\begin{figure}
\centering
\includegraphics[width=60mm, height=80mm, angle=-90]{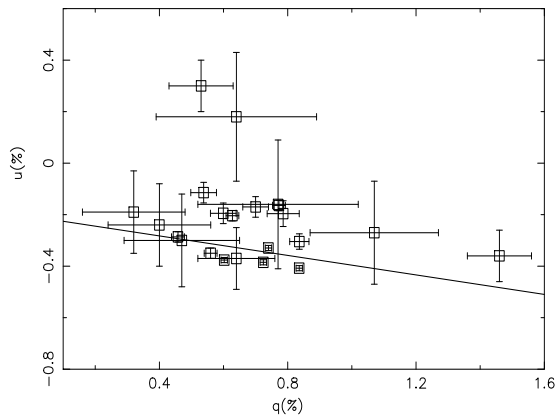}
\caption{(Q,U) data of HD~45677 measured in the {\it B}-band over the last 30 years. The data can be well fitted by a least-squares fit, suggesting a linear relationship between Q and U and constant PA. As the intrinsic polarisation angle does not change, the geometry of the circumstellar material remains constant over time.}
\label{F:HD45677_QU}
\end{figure}
In passing, we note that this value can not necessarily be compared to
the determination of the ISP by \cite{Coyne_Vrba:1976}, who derived a
Serkowski-law with the parameters p$_{max}$ = 0.65$\%$, $\theta$ =
170$^\circ$ at a $\lambda_{max}$ = 5000\AA$ $ using consistent
polarimetric measurements at short wavelengths (3650\AA). This value
was subsequently used by \cite{Schulte-Ladbeck_etal:1992} who found
that the polarisation increased by $>1\%$ from the {\it B}-band to the
{\it R}-band and a 90$^\circ$ flip in PA occured from the {\it UV} to
the red.  They also observe an enhancement in polarisation over
H$\alpha$ leading them to conclude that the Balmer line is formed
close to the star. It is important to note that Coyne $\&$ Vrba
calculated the ISP alone, while our value also takes into account the
polarisation due to circumsteller dust and therefore will be
different. Thus, the enhancements seen across H$\alpha$ and the
changes in polarisation observed by Schulte-Ladbeck et al. may still
be affected by foreground polarisation effects due to circumstellar
dust.  In our intrinsic spectrum, we find that depolarisations
occur across all of the Balmer lines, suggesting that the enhancements
observed were due to foreground polarisation effects and that the
Balmer lines are formed far from the electron-scattering region.

Both polarimetric and photometric variability can be seen in
HD~45677. These changes seem to be linked, with increasing
polarisation occuring during periods of decreasing visual brightness
and vice versa. We find that the {\it B}-band polarimetric data can be
well fitted using a least-squares fit, suggesting that the geometry of
the circumstellar material remains constant over time.  The gradient
of the fit is related to the intrinsic PA by Eqn. 1, giving an angle
of 175$^\circ \pm 1^\circ$, which is along the same plane as the
electron-scattering region as measured using the observed
line-effects.

The combination of photometric and polarimetric changes and the linear
distribution of the (Q,U) points suggests that the variations in
polarisation are due to changes in the dust polarisation component. De
Winter $\&$ van den Ancker (1997) explained that the photometric variations
may be caused by a wide binary or a blowout of material occuring in
the 1950s. Spectro-astrometric measurements of the star at H$\alpha$
by Baines et al. (2006) suggest the presence of a wide companion at a
PA of 150$^\circ$. However, a wide binary is unlikely to produce
significant variations in polarisation (1$\%$) over a period of 30
years (as the object would not intercept enough starlight for the
polarisation changes to be significant).

Alternatively, a blowout by the star may explain variations in the
{\it V}-band photometry and polarisation by, for example, the
creation and destruction of large grains. The very fact that we
observe any changes in polarisation implies the blowout was
aspherical.  Fig. \ref{F:HD45677_QU} shows that these changes occur
along a constant angle, therefore, this suggests that the blowout has
a flattened geometry at an intrinsic PA of 175$^\circ \pm
1^\circ$. The intrinsic polarisation angle associated with the line
effects across the Balmer lines showed that the aspherical ionised
region occurs at a PA of 164$^\circ \pm 3^\circ$. 

Monnier et al. (2006) interferometrically imaged HD~45677 in the
near-{\it IR} and proposed that the star is surrounded by an
elongated, skewed dust ring. They suggest that this geometry would
produce a net linear polarisation with a PA of 70$^\circ$. We propose
that the aspherical blowout occurs along the same plane as the ionised
region and, if the circumstellar environment is optically thin and
multiple scatterings are not important (see Vink et al. 2005), the
proposed angle of the blowout and the ionised region could be rotated
by 90$^\circ$ to define an angle on the sky of 85$^\circ \pm 1^\circ$
and 74$^\circ \pm 3^\circ$ respectively, which lies along the same
plane as the elongated dust ring proposed by Monnier et al. (2006).

\section{Conclusion}
We have presented new spectropolarimetric data combined with past
photometric and broad-band polarimetric observations on HD~45677. From
our spectropolarimetric data we have confirmed that the ionised region
is aspherical and that the Balmer lines in the {\it B} and {\it
R}-bands are formed far from the electron-scattering region.  We found
the object experiences variations in both {\it V}-band photometry
and {\it B} and {\it R}-band polarimetry and that these changes
are linked. We suggest that the results may be explained by a blowout
of material by the star occuring in the 1950s. In observing these
polarisation changes and from the (Q,U) data we conclude that this
blowout occured aspherically along the same plane as the ionised region.

\section*{Acknowledgements:}
We wish to thank the staff at the WHT for their assistance during the
observing runs. This work has made extensive use of the Starlink
software suite, in particular the packages FIGARO, TSP and POLMAP. We
acknowledge with thanks the variable star observations from the AAVSO
International Database contributed by observers worldwide and used in
this research. Finally, we thank the referee, Richard Ignace, for his
insightful comments that helped improve the presentation of the paper.

\bibliography{mnemonic,Patel_hd45677_ref}
\nocite{*}
\bibliographystyle{mn2e}
\bsp
\label{last page}
\end{document}